\documentclass[11pt]{article}
\usepackage{txfonts}
\usepackage{geometry}
\usepackage[utf8]{inputenc}
\usepackage{enumitem,amssymb}
\usepackage{ragged2e}
\usepackage[font=scriptsize,labelfont=bf]{caption}
\usepackage{pifont}
\usepackage{setspace}
\usepackage{eso-pic,graphicx}
\usepackage{tikz}
\usepackage{atbegshi}
\usepackage{url}
\usepackage{xcolor}    
\usepackage{multicol}
\usepackage[english]{babel}
\usepackage{authblk}
\definecolor{myblue}{RGB}{0,121,194}
\usepackage[colorlinks=true,
            linkcolor=myblue,
            urlcolor=myblue,
            citecolor=myblue]{hyperref}
\usepackage[capitalise,nameinlink]{cleveref}

\usepackage[
    style=numeric-comp,    
    sorting=none,     
    maxbibnames=3,     
    giveninits=true
]{biblatex}
\DeclareNameAlias{default}{family-given}
\DeclareNameAlias{author}{family-given}
\DeclareNameAlias{given}{initials}
\AtEveryBibitem{%
  \clearfield{title}%
  \clearfield{month}%
  \clearfield{day}%
  \clearfield{volume}%
  \clearfield{number}%
  \clearfield{pages}%
  \clearfield{doi}%
  \clearfield{issn}%
  \clearfield{isbn}%
  \clearfield{note}%
  \clearname{editor}%
}
\AtEveryBibitem{%
  \clearfield{journaltitle}%
  \clearfield{journal}%
  \clearfield{shortjournal}%
}
\renewbibmacro*{in:}{}
\DeclareFieldFormat{eprint}{\href{https://arxiv.org/abs/#1}{\textcolor{myblue}{arXiv}}}
\renewbibmacro*{eprint+url}{%
  \iffieldundef{eprint}{}{\printfield{eprint}}%
} 
\AtEveryBibitem{%
  \clearfield{month}%
  \clearfield{day}%
  \clearfield{volume}%
  \clearfield{number}%
  \clearfield{pages}%
  \clearfield{doi}%
  \clearfield{issn}%
  \clearfield{isbn}%
  \clearfield{note}%
  \clearfield{eid}%
  \clearfield{eprintclass}%
}
\DeclareFieldFormat{url}{\href{#1}{\textcolor{blue}{arXiv}}}
\renewbibmacro*{url+urldate}{%
  \iffieldundef{url}{}{%
    \printfield{url}%
  }%
}

\begin{document}

\selectlanguage{english}

\raggedright
\Large
ESO Expanding Horizons  \linebreak
\small
Transforming Astronomy in the 2040s \linebreak
Call for White Papers
\normalsize
\vspace{2.cm}

\begin{spacing}{1.6}
\textbf{\fontsize{17pt}{40pt}\selectfont
Kinematic lensing with high-resolution spectroscopic surveys}

{\fontsize{14pt}{20pt}\selectfont
A unique opportunity for transformative cosmology at high redshifts in the 2040s}
\end{spacing}

\vspace{0.5cm}
\textbf{Authors:} Stefano Camera,$^{1,2,3}$
Martin Kilbinger,$^4$ Jean-Paul Kneib,$^5$ Ofer Lahav,$^6$
Giovanni Aric\`o,$^7$ Sofia Contarini,$^8$ Giulia Degni,$^9$ Antonio Farina,$^{10}$ Massimo Guidi,$^{11,12}$ Vanshika Kansal,$^{13,14}$ Federico Marulli,$^{11}$ Alejandra Melo,$^{15}$ and Simone Sartori$^9$

\vspace{0.3cm}
\textbf{Contacts:} \href{mailto:stefano.camera@unito.it}{stefano.camera@unito.it}

\vspace{0.3cm}
\textbf{Affiliations:} \\
{\footnotesize
$^1$ Dipartimento di Fisica, Universit\`a degli Studi di Torino, 10125 Torino, Italy\\
$^2$ INFN -- Istituto Nazionale di Fisica Nucleare, Sezione di Torino, 10125 Torino, Italy\\
$^3$ INAF -- Istituto Nazionale di Astrofisica, Osservatorio Astrofisico di Torino, 10025 Pino Torinese, Italy\\
$^4$ Universit\'e Paris-Saclay, Universit\'e Paris Cit\'e, CEA, CNRS, AIM, 91191 Gif-sur-Yvette, France\\
$^5$ EPFL -- \'Ecole Polytechnique F\'ed\'erale de Lausanne, 1015 Lausanne, Switzerland\\
$^6$ Department of Physics and Astronomy, University College London, London WC1E 6BT, UK\\
$^7$ INFN -- Istituto Nazionale di Fisica Nucleare, Sezione di Bologna, 40127 Bologna, Italy\\
$^8$ Max Planck Institute for Extraterrestrial Physics, Giessenbachstrasse 1, 85748 Garching, Germany\\
$^9$ Aix Marseille Universit\'e, CNRS/IN2P3, CPPM, Marseille, France\\
$^{10}$ INAF -- Istituto Nazionale di Astrofisica, Osservatorio Astronomico di Brera, 20122 Milano, Italy \\
$^{11}$ Dipartimento di Fisica e Astronomia ``Augusto Righi'', Alma Mater Studiorum Universit\`a di Bologna, 40129 Bologna, Italy\\
$^{12}$ INAF -- Istituto Nazionale di Astrofisica, Osservatorio di Astrofisica e Scienza dello Spazio di Bologna, 40129 Bologna, Italy\\
$^{13}$ Centre for Astrophysics and Supercomputing, Swinburne University of Technology, Victoria 3122, Australia\\
$^{14}$ ARC Centre of Excellence for Dark Matter Particle Physics, Victoria 3122, Australia\\
$^{16}$ ESO -- European Southern Observatory, 85748 Garching-bei-Mu\"nchen, Germany\\
}

\pagenumbering{gobble} 

\pagebreak
\pagenumbering{roman}

\justifying

\begin{figure}
    \centering
    \includegraphics[width=\textwidth]{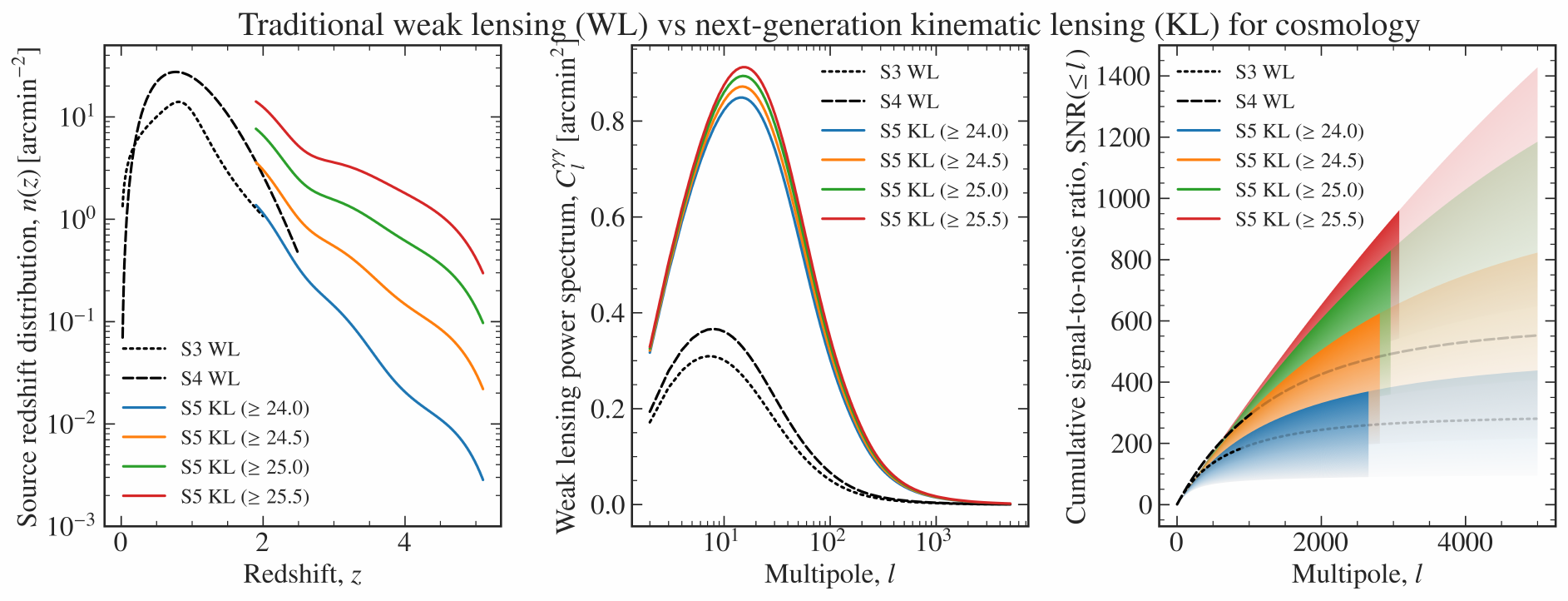}
    \caption{
    \textit{Left panel:} Galaxy angular densities for cosmic shear studies as a function of redshift. Black curves show analytical models for the redshift distribution of galaxies from photometric surveys of the previous generation (S3, dotted) or of the current one (S4, dashed), whereas coloured curves illustrate the expectation for S5 spectroscopic surveys (for more conservative to more optimistic magnitude thresholds). \textit{Middle panel:} Expected cosmic shear power spectrum as a function of angular multipole for the surveys of the right panel. \textit{Right panel:} Signal-to-noise ratio of the cosmic shear power spectrum, cumulative over the angular multipole \(l\), for the galaxy samples under consideration. The coloured areas span the cases where only \(0.5\%\) to \(5\%\) (from lighter to darker) of the galaxies in the left panel will be suitable for KL shear estimates. The different extension of the non-transparent curves/areas marks the range of multipoles corresponding to strictly linear scales.}
    \label{fig:placeholder}
\end{figure}

\paragraph*{The large-scale structure: the next frontier of cosmology.}
In the past decades, cosmological measurements have reached the level of percent accuracy on model parameters, establishing the \(\Lambda\)-cold dark matter (\(\Lambda\)CDM) model as the standard model of cosmology. 

However, from the theoretical point of view, we still lack a thorough understanding of its basic ingredients and the validity of its underlying assumptions: What is dark matter? What is driving the current accelerated expansion of the Universe---a cosmological constant \(\Lambda\) or some exotic dark energy component? Is general relativity valid on cosmological scales? Moreover, so-called `tensions', e.g.\ on the value of the Hubble constant \(H_0\), are currently raising increasing interest in the community, for they can point at cracks in our theoretical framework and lead to new physics.

In this, the large-scale structure (LSS) of the Universe is arguably the next frontier for cosmological investigation. If we quantify the information content by the number of Fourier modes accessible to measurements, it is estimated that the cosmic LSS contains \(10^6\)--\(10^8\) more modes than the cosmic microwave background (CMB) \cite{2016JCAP...06..046S,2016PhRvD..93h3510M}. It is not surprising, then, the extraordinary effort by the international community to probe the LSS up to an unprecedented extent, with instrumental observational campaigns \cite{2019BAAS...51c.341D,2019BAAS...51c.310P,2019BAAS...51c.101K}.\footnote{See also e.g.\ \texttt{\url{https://www.astronet-eu.org}}.}

Here, we argue that the proposed next generation (Stage V, or S5) of spectroscopic instruments for cosmology will be able to perform at high redshifts measurements of weak gravitational lensing via a novel technique called `kinematic lensing' (KL), thus enabling the unleashing of the full power of galaxy clustering-cosmic shear joint analyses in the next era of cosmological surveys of the LSS (see \cref{fig:placeholder}).

\paragraph*{The state of the art: galaxy clustering and cosmic shear.}
Traditionally, studies of the cosmic LSS have focussed on the clustering of galaxies. It involves scrutinising the spatial distribution of galaxies over large swathes of cosmic volumes, assuming galaxies as discrete tracers of the underlying continuous matter distribution \cite{2018arXiv180310814G,2018PhR...733....1D}. Galaxy clustering has proven most informative a cosmological probe, providing us with constraints on cosmological parameters comparable to those obtained with CMB analyses \cite{2022AJ....164..207D,2025arXiv250314745D}.

Over the past decades, another observable of the LSS has emerged: the weak gravitational lensing effect of shear. Photons from distant galaxies get scattered travelling towards us, due to intervening matter along the line of sight, resulting in distorted galaxy images \cite{2001PhR...340..291B}. In the regime of weak lensing, where the distortions are caused not by a single/few large matter fluctuation(s) aligned between the observer and the source, but by the entire inhomogeneous distribution of the cosmic LSS, lensing induces a correlation on the orientation of galaxy shapes---dubbed cosmic shear---that is detectable at a statistical level \cite{2015RPPh...78h6901K}.

What is most important is that galaxy clustering and cosmic shear are highly complementary. The former grants us direct access to the three-dimensional distribution of matter in the Universe and allows us to track the evolution of cosmic structures across time. However, it is a biased tracer of the LSS, as ordinary matter is but one sixth of the total (ordinary+dark) matter in the Universe, and galaxies themselves constitute a fraction of it. The latter, shear, is instead an unbiased tracer of the total matter distribution, but it is sensitive only to the integrated matter along the line of sight. Their combination, on the other hand, is less sensitive to the theoretical systematics due to galaxy bias and still allows for a tomographic study of the LSS.

A further added value of the combination of galaxy clustering and cosmic shear is that the former (assuming the validity of Euler's equation) is sensitive to the Newtonian gravitational potential---as galaxies `fall' in the potential wells of the LSS; whereas the latter is sourced by distortions of spacetime itself---thus being a truly relativistic probe. Hence, their combination
opens up transformational opportunities to distinguish between a cosmological constant, dynamical dark energy, and modified gravity \cite{2004PhRvD..70d3009H,2004ApJ...600...17B,2020A&A...642A.191E,2023PDU....3901151C}.

For all these reasons, galaxy clustering and cosmic shear represent the primary targets of state-of-the-art observational campaigns for cosmology, such as the surveys carried out---as of now or presently---by Stage-IV (S4) cosmological surveys, like DESI, 4MOST, \textit{Euclid}, \textit{Roman}, Rubin-LSST, and PFS.

\paragraph*{A science case for future spectroscopic galaxy surveys: kinematic lensing.}
The astrophysics and cosmology communities are aware of the challenges that current surveys are facing and are already planning future experiments and facilities to advance the state of the art ever forward. Of particular interest in this context are proposed spectroscopic instruments of the new, S5 generation, such as the MegaMapper \cite{2022MegaMapper_} or the Wide-field Spectroscopic Telescope \cite[WST,][]{2024WST_}. They aim to push to currently inaccessible redshifts the detection of galaxy positions with spectroscopic accuracy over large sky areas. By targetting Lyman-break galaxies (LBGs), made available thanks to deep \(u\)-, \(g\)-, and \(r\)-band photometry from the Vera C.\ Rubin Observatory, S5 surveys plan to collect a large sample of spectroscopic redshifts for galaxies in as deep a redshift range as \(z\in[2,5]\), and over almost half of the entire sky.

Such an unprecedented depth and corresponding gargantuan cosmic volumes will allow cosmologists to probe the growth of structures deep in the matter-dominated era and across the onset of dark-energy domination, offering a unique opportunity to unveil the mystery of cosmic acceleration.
It is then apparent that complementing such high-redshift galaxy clustering analyses with cosmic shear will materialise invaluable opportunities for cosmology. And it is in this context that KL may play an unprecedented role. 

KL is a novel technique that relies on modelling sheared velocity fields to recover the amount of deformation that galaxy images underwent by weak gravitational lensing by the intervening LSS \cite{2002ApJ...570L..51B,2006ApJ...650L..21M}.
%
Furthermore, KL offers an independent, pointwise measurement of the gravitational lensing effects of convergence and shear on a galaxy-by-galaxy basis. In other words, KL does not necessitate of large sample averages and, thus, is achievable with a smaller number of galaxies than required by traditional photometric imaging techniques.

\paragraph*{Our proposal: cosmology with a high-redshift kinematic lensing survey.}
As a proof of concept, we compare the forecast signal-to-noise ratio (SNR) for cosmic shear from proposed S5 galaxy surveys to that of the previous and current generations of shear surveys (respectively labelled here S3 and S4). We start by computing the expected angular redshift distribution of galaxies, \(n(z)\), for S5 surveys like WST or the MegaMapper, following \cite{2019Wilson&White,2021JCAP...12..009M,2025Rossiter_}. The left panel of \cref{fig:placeholder} shows the outcome (various colours for different limiting magnitudes), alongside analytical models for \(n(z)\) of S3 surveys like DES or CFHTLenS (black, dotted) or S4 experiments like \textit{Euclid} or Rubin-LSST (black, dashed).

Cosmic shear is an integrated effect, in the sense that the more distant a galaxy is, the more its image is going to be distorted, as photons emitted by it will have to cross an ever larger portion of LSS before reaching the observer. Therefore, the shear signal of galaxies at such high redshifts as those detected by the S5 spectroscopic instruments is expected to be significantly large. This can be easily appreciated in the middle panel of \cref{fig:placeholder}, where we present the anticipated cosmic shear signal, in the form of its harmonic-space power spectrum, \(C^{\gamma\gamma}_l\), as a function of the angular multipole \(l\). Cosmic shear from S5 instruments is, on average, five to eight times stronger than that measured by S3 and S4 surveys.

However, it is also clear from the left panel that the sheer (no pun intended) number of catalogued galaxies, proportional to the area subtended by the curves, will be smaller for S5 surveys---as expected when comparing spectroscopy versus multi-band imaging. This is incidentally why cosmic shear is traditionally performed on photometric galaxy catalogues: because averaging over a large number of galaxies is necessary to achieve enough SNR. In this respect, KL offers a major advantage, for the intrinsic scatter in the measurement of lensing shear per component can be more than an order of magnitude smaller than with traditional techniques \cite{2002ApJ...570L..51B,2021ApJ...922..116D,2023MNRAS.519.2535X,2024arXiv241000098H}. Considering e.g.\ the high-resolution spectroscopy of WST \cite{2024WST_}, we anticipate a shear rms per component of \(0.015\), versus \(0.2\)--\(0.3\) of traditional imaging \cite{2023MNRAS.519.2535X}.

Quantitatively, even if only a fraction of the galaxies detected by S5 experiments will have good enough resolved disk kinematics/velocity dispersion measurements to allow for KL, the reduced scatter will still allow for more than competitive performances. In the right panel of \cref{fig:placeholder} we show the SNR, cumulative over multipole, for all the surveys and configurations under consideration. Colour-shaded areas for KL S5 spectroscopic surveys, at different depths, span the very conservative to realistic cases in which only \(0.5\%\) (lighter) or a still cautious \(5\%\) (darker) of the galaxies will allow for KL shear estimates.

Despite the tiny angular number densities, the envisaged S5 SNR is on average twice as much that of S4 surveys, and several times that of S3 ones. For reference, the constraining power on \(S_8\propto\sigma_8\,\sqrt{\Omega_{\rm m}}\) (a parameter summarising the strength of clustering, \(\sigma_8\), and the present-day matter abundance, \(\Omega_{\rm m}\)) shrank by \({\sim}5\) from S3 to S4 experiments. Moreover, at high redshift the growth of cosmic structures is more linear than at later times, which implies that for S5 surveys a larger number of angular modes will be captured by the well-understood linear theory (non-transparent curves/areas in the right panel of \cref{fig:placeholder}), compared to S3 and S4 predecessors, thus strongly limiting the impact of theoretical systematics.

\paragraph*{The road ahead: technological challenges and scientific opportunities.}
KL is a promising yet novel technique, and at the level of cosmological inference it is but at its infancy. It requires high-quality spectroscopic measurements of galaxy rotation fields in addition to imaging, which is observationally expensive and technically demanding (e.g.\ high spectral resolution, good SNR, and overlapping imaging and spectroscopy), and it depends on accurate calibration of galaxy scaling relations (like Tully-Fisher) and/or control of astrophysical systematics that could bias the cosmological shear inference.

Nonetheless, most if not all of these requirements will be met by the next generation of spectroscopic instruments, in particular by those equipped with both high-resolution spectroscopy with multi-object spectrographs (MOSs) and panoramic, integral-field spectrographs (IFSs), like WST. This opens up transformational opportunities to extend to S5 surveys joint galaxy clustering-cosmic shear analyses, previously unthought-of at such high redshifts, and unattainable even with superb instruments like JWST or the ELT for their lack of wide-area coverage, essential for cosmology.


\printbibheading
\begin{multicols}{2}
\printbibliography[heading=none]
\end{multicols}

\end{document}